\documentclass[twocolumn,nofootinbib,showpacs,prd]{revtex4}

\usepackage{graphics,graphicx}
\usepackage{amsmath}
\usepackage{psfrag}
\usepackage{dcolumn}
\usepackage{psfrag}
\allowdisplaybreaks

\newcommand{\be}{\begin{equation}}
\newcommand{\ee}{\end{equation}}
\newcommand{\bea}{\begin{eqnarray}}
\newcommand{\eea}{\end{eqnarray}}
\def\no{\nonumber \\ & \quad}

\def\bes{\begin{subequations}}
\def\ens{\end{subequations}}
\def\bal{\begin{align}}
\def\eal{\end{align}}

\newcommand{\vek}[1]{\boldsymbol{#1}}

\begin{document}

\title{
Gravitational wave data analysis implications of TaylorEt inspiral 
approximants for ground-based detectors:
the non-spinning case   
}
\author{Sukanta Bose}
\email{sukanta@wsu.edu}
\affiliation{
Department of Physics \& Astronomy, Washington State University,
1245 Webster, Pullman, WA 99164-2814, U.S.A.
}
\altaffiliation{Currently at {\it Max-Planck-Institut f\"ur Gravitationsphysik,
Albert-Einstein-Institut, Callinstr. 38, 30167 Hannover, Germany}}
\author{Achamveedu Gopakumar}
\email{a.gopakumar@uni-jena.de}
\affiliation{Theoretisch-Physikalisches Institut, Friedrich-Schiller-Universit\"
at Jena, Max-Wien-Platz 1,07743 Jena, Germany}
\email{a.gopakumar@uni-jena.de}
\author{Manuel Tessmer}
\email{m.tessmer@uni-jena.de}
\affiliation{Theoretisch-Physikalisches Institut, Friedrich-Schiller-Universit\"
at Jena, Max-Wien-Platz 1,07743 Jena, Germany}

\date{July 15, 2008; LIGO Document Control Center No. LIGO-T080142-00-Z}

\begin{abstract}

A new family of restricted post-Newtonian-accurate 
waveforms, termed TaylorEt approximants,
was recently proposed for searching gravitational wave (GW) signals from 
inspiraling non-spinning compact binaries having arbitrary mass-ratios.
One of the attractive features exhibited by these waveforms is that as their 
reactive post-Newtonian (PN) order is increased, their
phase-evolution monotonically converges to that of waveforms produced
by numerical relativity.
The TaylorEt approximant is different from the usual post-Newtonian ones,
such as the TaylorT1, TaylorT4, and TaylorF2 approximants, in that
it gives equal emphasis to both the conservative and
the reactive parts of GW phase evolution.
However, for the latter set of extensively employed PN-accurate inspiral templates
the conservative phase evolution is somewhat
dwarfed by the reactive part.
We perform detailed fitting factor (FF) studies to
probe if the TaylorEt (3.5PN) signals 
for non-spinning comparable mass compact  binaries
can be effectually and faithfully
searched with TaylorT1, TaylorT4, and TaylorF2 (3.5PN)
templates in LIGO, Advanced LIGO, and Virgo interferometers.
We observe that a good fraction of the templates, which
by choice are from TaylorT1, TaylorT4, and TaylorF2 (3.5PN)
families, have ${\rm FF} ~{}^{\Large <}_{\sim}~ 0.97$ 
and substantial biases for the estimated total-mass
against the fiducial TaylorEt (3.5PN) signals for equal-mass systems. Both 
these observations can bear on the detectability of a signal.
TaylorEt (3.5PN) signals with mass-ratios of a third or a quarter
yield high FFs against those same template banks, but at the expense of inviting
large systematic errors in the estimated values of their
total mass and symmetric mass-ratio.
In general, the aforementioned templates are
found to be increasingly {\it unfaithful} with respect to a TaylorEt
signal as one increases the total mass of the inspiraling system.
We find that one way of 
improving the FF values is to
allow the templates to have unphysical mass-ratios. However, this
may result in a higher noise background and, therefore, reduce the detection 
confidence.
We also observe that 
the amount of bias in the estimated mass varies with the 
(noise power spectral density of the) detector. This can be of some concern 
for multi-detector searches, which check for consistency in the estimated 
masses of concurrent triggers in their data.
However, by modeling this variation it is possible
to mitigate its effect on the detection efficiency.

\end{abstract}


\maketitle

\section{Introduction}

 Stellar mass compact binaries, involving black holes and neutron stars, are 
the most promising sources 
of gravitational radiation for the operational and planned ground-based  laser interferometric 
gravitational wave (GW) detectors. Gravitational wave signals from inspiraling compact binaries are 
being searched in the detector data by using {\it matched filtering} 
\cite{Helstrom} with several 
types of theoretically modeled inspiral templates \cite{Blanchet:1995ez,Damour:2000zb}. A good resource for computing these templates in software, which is being actively used by the LIGO Scientific Collaboration (LSC) \cite{LSC} and the Virgo Collaboration \cite{VirgoC} for analyzing detector data, is 
the LSC Algorithm Library (LAL) \cite{LAL}.

As such, the construction of these
search templates requires {\it two} crucial inputs from 
the post-Newtonian (PN) approximation to general relativity, appropriate for 
describing the dynamics of a compact binary during its inspiral phase.
These are
the 3PN accurate dynamical (orbital) energy ${\cal E}(x)$
and the 3.5PN accurate expression for the GW luminosity
$ {\cal L }(x) $ \cite{PN_results}, both of which are 
usually expressed as PN series in the gauge invariant quantity $x\equiv (G\,m\, \omega/c^3)^{2/3}$,
where $m$ and $\omega$ are the total mass and the orbital angular frequency of the binary, 
and the familiar symbols $G$ and $c$ denote the universal gravitational constant and the speed of light 
in vacuum, respectively.
Recall that the 3PN accurate expression for ${\cal E}(x)$ provides
corrections to the Newtonian orbital energy to the order 
of $(v/c)^6$, where $v$ is the orbital speed. 
Further, the currently employed search templates only require the Newtonian contributions to
the amplitude of GW polarizations,
$h_{+} (t)$ and $h_{\times}(t)$.
However, expressions for $h_{+} (t)$ and $h_{\times}(t)$ that include the
3PN amplitude corrections are available in Ref.~\cite{BFIS} and are being used to
develop amplitude corrected templates for GW inspiral searches.

    With the help of the two aforesaid PN-accurate inputs, one can construct two 
distinct classes of inspiral GW templates.
The templates belonging to the first category require  
\be\label{phiEvolution}
\frac{d \phi (t)}{dt} = \omega (t)  \equiv \frac{c^3}{G\,m}\, x^{3/2}\,
\ee
and PN-accurate prescriptions for the reactive evolution of $x(t)$.
Such templates are usually referred to as {\em adiabatic} inspiral templates,
and all PN-accurate inspiral templates that LAL employs are of this type.
In this paper, we consider from this class time-domain templates of the
TaylorT1 \cite{PN_results} and the Numerical Relativity (NR) inspired TaylorT4 \cite{CC07}
families and frequency-domain templates of the TaylorF2 family \cite{AISS}.
For all three families, we incorporate radiation reaction effects to the 
(relative) 3.5PN order (see Eqs.~(\ref{EqP1}), (\ref{EqP3}), and (\ref{EqP4}) 
below). Due to their use of the $x$-based phase evolution expression 
in Eq. (\ref{phiEvolution}),
it may be argued that these templates
model GWs from compact binaries inspiraling under PN-accurate radiation reaction
along {\it exact} circular orbits \cite{AG07}.

  A new class of inspiral approximants introduced in Ref.~\cite{AG07}, 
termed as TaylorEt,
requires PN expansion for $d\phi/dt$ in terms of the orbital binding 
{\em Energy} to derive the {\em temporal} GW phase evolution.
This alternative phasing prescription models GWs from compact binaries
inspiraling under PN-accurate reactive dynamics along {\it PN-accurate 
circular orbits} \cite{AG07}. In other words, the TaylorEt approximant 
{\it explicitly} incorporates the secular contributions to GW phase evolution 
appearing at the 1PN, 2PN, and 3PN orders. Contrastingly, in the case of  
$x$-based adiabatic inspiral templates and due to the use of 
Eq. (\ref{phiEvolution}),
the above mentioned conservative (and secular) contributions to
the GW phase evolution do not appear before the radiation reaction kicks in at the
absolute 2.5PN order.  It should be noted that the cost of computing
TaylorEt templates is comparable to that of TaylorT1/T4 templates.

In this paper, we study how effectively and faithfully 
the TaylorT1, TaylorT4, and TaylorF2 inspiral templates, at 3.5PN order,
can capture a GW signal modeled using the TaylorEt approximant of the same order.
The main motivation for using the 
latter as the fiducial signal
originates from the observation that the TaylorEt approximant is an appropriate
zero-eccentricity limit of GW phasing for compact binaries inspiraling along PN-accurate eccentric 
orbits \cite{AG07}.
We quantify our results by computing fitting factors (FF)
following prescriptions detailed in Refs.~\cite{A95,DIS}, 
and inherent systematic errors
in the estimated value of $m$ and the symmetric mass-ratio, $\eta$, for the 
 various search templates,
relevant for the initial LIGO (heretofore referred to as ``LIGO''), 
Advanced LIGO (or ``AdLIGO'') and Virgo detectors. 
We conclude that it is desirable to incorporate the TaylorEt approximant at 3.5PN order
into LAL to minimize possible loss of inspiral events.
Further, one might also view this work as an exercise in
assessing the effects of using inspiral templates from different
representations on a GW signal's detectability and parameter
estimation in earth-based detectors. Similar assessments of
systematic errors on GW searches in LISA and Virgo
were made by comparing inspiral
templates of different PN orders from the same representation in
Refs.~\cite{CV07} and ~\cite{PC01}, respectively.

The plan of the paper is as follows. In the next section, Sec.~\ref{approximants}, we provide 
explicit PN-accurate equations
required for constructing TaylorT1, TaylorT4, TaylorF2 and TaylorEt templates
having 3.5PN accurate reactive evolution. Section~\ref{FF} explains how we perform
our FF computations and tabulates their values, along with the associated systematic
errors in $m$ and $\eta$,
for our different templates. We briefly discuss 
the implications of these results on the on-going searches in 
real interferometric data.
We conclude in Sec.~\ref{conclusions} by providing a brief summary and future directions. 

\section{Phasing formulae for various inspiral templates}
\label{approximants}

 The PN approximation to general relativity is expected to describe accurately 
the adiabatic inspiral phase of a comparable mass compact binary \cite{CC07}. 
During this phase, the change in the orbital frequency over one orbit 
may be considered to be tiny compared to the mean orbital frequency itself.
For compact binaries, having negligible eccentricities, the 
adiabatic orbital phase evolution can be 
accurately described with the help of 
3PN and 3.5PN accurate expressions for the 
orbital energy and the GW luminosity, respectively, available in Refs.~\cite{PN_results}.
 While employing $x$ as a PN expansion parameter, 
there exist several prescriptions to compute the adiabatic GW phase evolution.
Each prescription, termed a PN
approximant, provides a slightly different GW phase evolution and, 
correspondingly, a different inspiral template family.
Following Ref.~\cite{LAL}, we first list 
the equations describing the TaylorT1 and the TaylorF2 approximants, 
which are regularly employed by various GW data analysis groups.
The time-domain TaylorT1 approximant is given by
\begin{subequations}
\label{EqP1}
\begin{align}
\label{EqP1a}
h(t) & \propto x\, \cos 2\,\phi(t) 
\,,\\
\label{EqP1b}
\frac{d \phi (t)}{dt} & = \omega (t)  \equiv \frac{c^3}{G\,m}\, x^{3/2}\,,\\
\frac{d\,x(t)}{dt} &=  -\frac{{\cal L}( x)}{ \left( d {\cal E} / d x
\right)}\,,
\label{EqP1c}
\end{align}
\end{subequations}
where the proportionality constant in Eq. (\ref{EqP1a}) 
may be set to unity for our analysis.
To construct the TaylorT1 3.5PN order adiabatic inspiral templates, one needs to use
3.5PN accurate ${\cal L}(x)$ and 3PN accurate ${\cal E}(x)$,
respectively. The explicit expressions for these quantities, extracted from Refs.~\cite{PN_results},
read
\begin{subequations}
\label{EqP2}
\begin{align}
\label{EqP2a}
{\cal L}(x) &= \frac{32\,\eta^2\,c^5}{5\,G}\, x^{5}\,
\biggl \{
1
- \biggl [ {\frac {1247}{336}}+{\frac {35}{12}}\,\eta \biggr ] x
+4\,\pi \,{x}^{3/2}
\no
- \biggl [ {\frac {44711}{9072}}
-{\frac { 9271}{504}}\,\eta
-{\frac {65}{18}}\,{\eta}^{2} \biggr ] {x}^{2}
- \biggl [
{\frac {8191}{ 672}} 
\no
+{\frac {583}{24}}\,\eta
 \biggr ]\, \pi\, {x}^{5/2}
+ \biggl [
{\frac {6643739519}{69854400}}
+\frac{ 16\, {\pi }^{2}}{3} 
\no
-{\frac {1712}{ 105}}\,\gamma
- \left({\frac {134543}{7776}}-{\frac {41}{48}}\,{\pi }
^{2} \right) \eta
-{\frac {94403}{3024}}\,{\eta}^{2}
\no
-{\frac {775}{324}}
\,{\eta}^{3}-{\frac {1712}{105}}\,\ln  \left( 4\,\sqrt {x} \right) 
\biggr ]
 {x}^{3}
- \biggl [ 
{\frac {16285}{504}}\, 
\no
-{\frac {214745}{ 1728}}\,\eta
- {\frac {193385}{3024}}\,{\eta}^{2}
\biggr ]\, \pi\,
{x}^{7/2}
\biggr \}
\,,\\
{\cal E}(x) &= -\frac{\eta\, m\, c^2}{2}\,x
\biggl \{
1
- \frac{1}{12} \biggl [ 9 +  \eta \biggr ] x
- 
\biggl [
{\frac {27}{8}} 
-{ \frac {19}{8}}\,\eta
\no
+\frac{1}{24}\,{\eta}^{2}
\biggr ]{x}^{2}
- \biggl [  
{\frac {675}{64}}+{\frac {35}{5184}}\,{\eta}^{3}+{\frac {
155}{96}}\,{\eta}^{2}
\no
+ \left( {\frac {205}{96}}\,{\pi}^{2}-{\frac {
34445}{576}} \right) \eta
\biggr ]
{x}^{3}
\biggr \}
\,,
\label{EqP2b}
\end{align}
\end{subequations}
where  $\gamma$ is the Euler constant and 
$\eta \equiv \mu/m$, with $\mu$ being the reduced mass of the binary.

 The frequency-domain TaylorF2 approximant at 3.5PN order, extracted from Ref.~\cite{AISS}, reads
\begin{subequations}
\label{EqP3}
\begin{align}
\label{EqP3a}
\tilde h(f) & \propto  f^{-7/6}\, e^{i\, \psi(f)}\,,\\
\psi(f) &= 2\, \pi\, f\, t_c - \phi_c -\frac{\pi}{4}  \no
 + \frac{3}{128\, \eta\, (v/c)^5}
\sum_{k=0}^{k=7} \alpha_k\, \left(\frac{v}{c}\right)^{k}\,,   
\label{EqP3b}
\end{align}
\end{subequations}
where $ v = ( G\pi\, m\, f / c^3)^{1/3}$, and $t_c$ and $\phi_c$ are the fiducial time 
and phase of coalescence, respectively.
The explicit expressions for the PN coefficients $\alpha_k$ are
\begin{subequations}
\begin{align}
\alpha_0&=1,\\
\alpha_1&=0,\\
\alpha_2&=\frac{20}{9}\,\left( \frac{743}{336} + \frac{11}{4}\eta
\right),\label{Eq:alpha2}\\
\alpha_{3}&= -16\pi,\label{Eq:alpha3}\\
\alpha_4&=10\,\left( \frac{3058673}{1016064} + \frac{5429\,
}{1008}\,\eta + \frac{617}{144}\,\eta^2 \right),\\
\alpha_5&=\pi\biggl\{\frac{38645 }{756}+ \frac{38645 }{252}\,
\log \left(\frac{v}{v_{\rm lso}}\right) 
\no
- {65\over9}\eta\left[1  +
3\log \left(\frac{v}{v_{\rm lso}}\right)\right]\biggr\},\\
\alpha_{6}&=\left(\frac{11583231236531}{4694215680} - \frac{640\,{\pi
}^2}{3} -
\frac{6848\,\gamma }{21}\right)
\no
-\eta \,\biggl ( \frac{15737765\,635}{3048192} - \frac{2255\,{\pi }^2}{12}
\biggr )
\no
+{76055\over 1728}\eta^2-{127825\over 1296}\eta^3-{6848\over 21}
\log\left(4\;{v}\right),\\
\alpha_7 &=\left(\frac{77096675 }{254016} + \frac{378515
}{1512}\,\eta - \frac{74045}{756}\,\eta^2\right)\,\pi\,,
\end{align}
\end{subequations}
where $v_{\rm lso}$ is the speed at the last stable orbit,
which we take to be at
$6\, Gm/c^2$. 

  Recently, Ref.~\cite{CC07} introduced 
another Taylor approximant, termed TaylorT4.
This approximant is obtained by Taylor expanding in $x$ the right-hand side
of Eq.~(\ref{EqP1c}) for $dx/dt$
and truncating it at the appropriate reactive PN order.
This approximant at 3.5PN order has an interesting (and accidental) property that was 
discovered due to the recent advances in Numerical Relativity (NR) involving 
coalescing binary black holes \cite{FP}. 
It was observed in 
Ref.~\cite{CC07} that the NR-based GW phase evolution for an equal-mass binary black hole 
agrees quite well with its counterpart in TaylorT4 approximant at 3.5PN order.
Specifically,  Ref.~\cite{CC07} observed that the accumulated GW phase difference between 
TaylorT4 waveforms at 3.5PN order and NR waveforms agrees within  0.06 radians
over 30 wave cycles and matched at $x \sim 0.215$.
The time-domain TaylorT4 approximant at 3.5PN order is specified by
\begin{subequations}
\label{EqP4}
\begin{align}
\label{EqP4a}
h(t) & \propto x\, \cos 2\,\phi(t) 
\,,\\
\label{EqP4b}
\frac{d \phi (t)}{dt} & = \omega (t) \equiv \frac{c^3}{G\,m}\, x^{3/2}\, ,\\
\label{EqP4c}
\frac{d\,x(t)}{dt} &=  \frac{c^3}{G\,m}\, \frac{64\,\eta}{5}\, x^5
\biggl \{
1
- \left( {\frac {743}{336}}+ \frac{11}{4}\,\eta \right) x
+4\,\pi\,{x}^{3/2}
\no
+ \left( {\frac {34103}{18144}}+{\frac {13661}{2016}}\,\eta+{\frac {59}
{18}}\,{\eta}^{2} \right) {x}^{2}
- \biggl [ {
\frac {4159}{672}}\,
\no
+{\frac {189}{8}}\,\eta \biggr]\,\pi\, {x}^{5/2}
+ \biggl [ 
{\frac {16447322263}{139708800}}
-{\frac {1712}{105}}\, \gamma 
\no
+\frac{16\,{\pi}^{2}}{3}-{\frac {3424}{105}}\,\ln  \left( 2 \right) -{
\frac {856}{105}}\,\ln  \left( x \right)  
- \biggl ( {\frac {56198689}{217728}}
\no
-{\frac {451}{48}}\,{\pi}^{2} 
\biggr )
\eta
+{\frac {541}{896}}\,{\eta}^{2}
-{ \frac {5605}{2592}}\,{\eta}^{3}
\biggr ]
 {x}^{3}
- \biggl [ 
 {\frac {4415}{4032}}\,
\no
-{\frac {358675}{6048}}\,\eta
-{\frac {91495}{1512}}\,{\eta}^{2} 
\biggr ]\,\pi\,
 {x}^{7/2}
\biggr \}
\,.
\end{align}
\end{subequations}
It should be noted that the TaylorF2 waveform in Eqs. (\ref{EqP3}) is
the Fourier transform of $h(t)$, given by Eqs.~(\ref{EqP4}) above,
computed with the help of the 
stationary phase approximation \cite{BO}; we speculate 
that this is the reason
that the TaylorT4 approximant is not directly employed in LAL
(which already uses the TaylorF2 approximant).

 A close inspection of various time-domain 
adiabatic inspiral templates available in LAL
reveals that they all invoke Eq. (\ref{phiEvolution}).
These template families are different from one another only in the manner 
in which they incorporate the reactive evolution of $x(t)$.
For example, PadeT1 time-domain inspiral templates are constructed by 
invoking a specific Pade resummation for the right-hand side of Eq.~(\ref{EqP4c}). 
Therefore, we may state that various $x$-based inspiral templates provide 
slightly different GW phase evolution
by perturbing a compact binary in an exact circular orbit, 
defined by Eq. (\ref{phiEvolution}),
by different prescriptions
for the reactive evolution of $x(t)$.
This is the main reason 
behind the observation that 
these templates
model GWs from compact binaries inspiraling under PN-accurate radiation reaction along
{\it exact} circular orbits.

   Interestingly, it is possible to construct, in a gauge-invariant manner, inspiral GW search templates
that do not require working in terms of the $x$ variable. 
The TaylorEt approximant
\cite{AG07}
employs the orbital binding energy in lieu of that variable
to describe PN-accurate adiabatic GW phase evolution.
Hence, it requires an appropriate PN expansion for  $d\phi/dt$ in terms of 
the orbital binding energy.
Accordingly, it can be argued that the TaylorEt approximant  
models GWs from compact binaries inspiraling under
PN-accurate radiation reaction along {\it PN-accurate circular orbits}.

        The TaylorEt approximant at 3.5PN order is defined by
\begin{subequations}
\label{EqP6}
\begin{align}
h(t) & \propto {\cal E}(t)\, \cos 2\,\phi(t)
\label{EqP6a}
\,,\\
\label{EqP6b}
\frac{d \phi (t)}{dt} &\equiv \omega (t) = \frac{c^3}{G\,m}\, \xi^{3/2}
\biggl \{
1
+ {\frac {1}{8}}
\left( {9}+\eta \right) \xi
+ \biggl [
{\frac {891}{128}} 
\no
-{\frac {201}{ 64}}\,\eta
+{\frac {11}{128}}\,{\eta}^{2} 
\biggr ]
{\xi}^{2}
+ \biggl [
{\frac {41445}{1024}}
- \biggl ( {\frac {309715}{3072}}
\no
-{\frac {205}
{64}}\,{\pi}^{2} \biggr) \eta 
+{\frac {1215}{1024}}\,{\eta}^{2}
+{\frac {45}{1024}}\,{\eta}^{3}
\biggr ]
 {\xi}^{3}
\biggr \}
\,,\\
\frac{d\,\xi (t)}{dt} &= {\frac {64}{5}}\,\eta\,{\xi}^{5}
\biggl \{
1
+ \left( {\frac {13}{336}}- \frac{5}{2}\,\eta \right) \xi
+4\,\pi\,{\xi}^{3/2}
\no
+ \left(
{\frac {117857}{18144}}
-{\frac {12017}{2016}}\,\eta
+\frac{5}{2}
\,{\eta}^{2} \right) {\xi}^{2}
+ 
\biggl [
{\frac {4913}{672}} 
\no
-{\frac {177}{8}}\,\eta
\biggr ]\, \pi\,
 {\xi}^{5/2}
+ 
\biggl [
{\frac {37999588601}{279417600}}
\no
-{\frac {1712}{105}}\,\ln  \left( 4\,\sqrt {\xi} 
\right) -{\frac {1712}{105}} \,\gamma 
+\frac{16\,{\pi}^{2}}{3}
\no
+ 
\biggl (
{\frac {369}{32}}\,{\pi}^{2}-{\frac {24861497}{72576}} 
\biggr )
 \eta
+{\frac {488849}{16128}}\,{\eta}^{2}
\no
-{\frac {85}{64}}\,{\eta}^{3}
\biggr ]
 {\xi}^{3}
+ 
\biggl [
{\frac {129817}{2304}}\, 
-{\frac {3207739}{48384}}\,\eta
\no
+ {\frac {613373}{12096}}\,{\eta}^{2}
\biggr ]\, \pi\,
 {\xi}^{7/2}
\biggr \}
\,,
\label{EqP6c}
\end{align}
\end{subequations}
where
$\xi = -{2\, \cal E}/\mu\,c^2$.
A close inspection of Eqs.~(\ref{EqP6}) reveals that the above inspiral $h(t)$ is obtained by
perturbing a compact binary in a 3PN accurate circular orbit, defined by 
Eq.~(\ref{EqP6b}), by radiation reaction effects at 3.5PN order, given by Eq.~(\ref{EqP6c}).
The explicit use of PN-accurate expression for $d \phi/dt$ allows us 
to state that the
TaylorEt approximants
model GWs from compact binaries inspiralling under PN accurate reactive dynamics
along PN accurate circular orbits.

Importantly, a recent study of the accumulated phase difference
between NR waveforms on the one hand and TaylorEt, TaylorT1, and TaylorT4 
waveforms on the other hand
reveals the following characteristics \cite{GHHB}.
In the interval $x \sim 0.127 $ to $x \sim  0.215 $ this difference for
the TaylorEt approximant at 3.5PN order is $\delta \phi \sim -1.18$ radians, 
which is more than what is found for the TaylorT1 and TaylorT4 counterparts 
(with $\delta \phi \sim 0.6$ and $0.06$ radians, respectively). 
However, significantly, TaylorEt is the only approximant studied so far that 
exhibits monotonic phase convergence with the NR waveforms when
its reactive PN order is increased \cite{GHHB}. Recall that sophisticated 
Pad\'{e} approximations are required to make
$x$-based Taylor approximants converge monotonically to the $h(t)$ obtained
from numerical relativity in the  $\eta = 0$ case \cite{DIS}. 
In the context of the
present paper, the analysis detailed in Ref.~\cite{GHHB} also suggests that 
TaylorEt approximant at 3.5PN order remains
fairly accurate in describing the inspiral $h(t)$ even near
the last stable orbit. These properties make it worthwhile 
to study the data analysis implications of TaylorEt
approximants.
  
Another motivation for using the TaylorEt approximant to model the 
expected inspiral GW signal is as follows: With the help of 
Refs.~\cite{AG07,DGI}, it can be argued that TaylorEt is an appropriate 
approximant resulting from the zero-eccentricity limit of GW phasing 
of compact binaries inspiraling along PN-accurate eccentric orbits.
By contrast, the construction of the usual {\em adiabatic} inspiral templates
requires redefining the right-hand side of Eq.~(\ref{EqP6b}) to be 
$c^3 x^{3/2}/ Gm$,
which can not be extended to yield GWs from precessing and inspiraling eccentric
binaries as obtained in Ref.~\cite{DGI}.
Therefore, the TaylorEt approximant can be expected to closely 
model GW signals from inspiraling compact binaries,
which realistically will not move along exactly circular orbits.
The above statements are based on Ref.~\cite{TG07} that, while restricting 
radiation reaction to the dominant quadrupole contributions, demonstrated
the undesirable consequences of redefining the 
right-hand side of
Eq.~(\ref{EqP6b}) at 2PN order to be $c^3 x^{3/2}/ Gm$.

 Currently, for the low-mass binary signal 
searches the LSC usually employs templates 
based on TaylorT$n$ (where $n$=1, 2, and 3) and TaylorF2 approximants 
\cite{LAL,Abbott:2007xi}.
Therefore, it is important to probe if some of these templates
can capture inspiral signals modeled on the TaylorEt approximant. 
This is what we pursue in the next section.
 
\section{Fitting Factors}

\label{FF}
Inspiraling compact binaries are the most promising sources of GWs for 
LIGO/Virgo. Detailed source population synthesis studies 
suggest that achieving an appreciable 
event rate, of at least a few compact binary coalescences per year,
is possible if one could hear sources in the far reaches of our local super-cluster and beyond \cite{Kalogera:2003tn,O'Shaughnessy:2005qs}.
Such an endeavor necessitates the ability to detect signals with relatively 
low signal-to-noise ratios (SNRs), even with second generation detectors, such 
as AdLIGO. Let the GW strain from a non-spinning compact
binary be denoted by $h(t;\vek{\lambda})$, where $\vek{\lambda}$ represents
the signal parameters, namely, $m$, $\eta$, $t_c$, and $\phi_c$,
or an alternative set of transformed coordinates
in that parameter space. If a detector's strain-data is denoted by $s(t)$ 
and its noise power-spectral-density (PSD) by $S_n(f)$, then the SNR when
filtering the data with template $h(t;\vek{\lambda'})$ is
\be\label{SNRDef}
{\rm SNR} = \frac{\langle s | h(\vek{\lambda'}) \rangle }
{\sqrt{ \langle h(\vek{\lambda'}) |h(\vek{\lambda'}) \rangle}}
\,,
\ee
where  $\vek{\lambda'}$ symbolizes the template parameters, which need not 
be the same as the parameters of a signal embedded in the data, and the
inner product $\langle a|b \rangle$ is defined as,
\be
\langle a|b \rangle \equiv 4\Re \int_0^{\infty} \frac{\tilde a^*(f) \tilde b(f)}{S_n(f)}\,df
\,.
\ee
Above, $\tilde{a}(f)$ is the Fourier transform of $a(t)$ and the asterisk denotes 
complex conjugation. Using Eq.~(\ref{SNRDef}), it can be shown that the quantity
\be\label{match}
{\rm M}\left( \vek{\lambda}, \vek{\lambda'} \right) = \frac{\langle g(\vek{\lambda}) | h(\vek{\lambda'}) \rangle }
{\sqrt{ \langle g(\vek{\lambda}) |g(\vek{\lambda}) \rangle \langle h(\vek{\lambda'}) |h(\vek{\lambda'}) \rangle}}
\,,
\ee
also known as the ``match'', is useful in describing how well two 
normalized waveforms, not necessarily from 
the same template family, overlap \cite{BO96}.

For the problem of detecting a GW inspiral signal, the prevailing 
sentiment in the community is that it is not as essential 
to search with a template bank that 
is an exact representation of the signal,
as it is to search with an approximate one that 
can filter the data in real-time, provided its 
expected maximal match with a signal from anywhere in the parameter space 
is above a desired threshold. 
In other words, it should be possible to obtain a sufficiently large `match' with 
a family of templates having $\vek{\lambda'} \neq \vek{\lambda}$.
 It is often stressed that this 
faithlessness of a template in accessing the signal parameters does not
concern the detection problem {\it per se}, but that it affects 
the parameter-estimation problem, 
which can be tackled {\it a posteriori}, i.e., after the transient signal
has been detected and localized in time. 
The effectiveness of a template family, say, $h(\vek{\lambda'})$, in detecting 
the target signal $g(\vek{\lambda})$ is
quantified by the fitting factor (FF) \cite{A95}
\be\label{FFDef}
{\rm FF}(\vek{\lambda}) = \max_{\vek{\lambda'}} \, {\rm M}\left( \vek{\lambda}, \vek{\lambda'} \right)\,.
\ee
If a template bank provides near-unity FF values for a given signal,
it is considered to be {\it effectual} in detecting it
\cite{DIS}.

Employing an approximate template bank results in a drop in event rate
by $(1 - {\rm FF}^3)$ for a homogeneous distribution of sources. This is
easily seen when one realizes that the FF 
is a measure of the fractional loss in SNR (which scales inversely with 
source distance) stemming from using such a bank. So, e.g., a FF of 90\% 
results in a 27\% loss in event rate in any given detector. The
expected rate in LIGO or Enhanced LIGO, which is a proposed upgrade
in sensitivity of LIGO by roughly a factor of two while making minimal changes
to the shape of the LIGO noise PSD \cite{ELIGOFFs}, is very low,
i.e., realistically, less than one event in a few years. Therefore, a
FF of 90\% can potentially subvert a detection in the era of first-generation
detectors. This is why a FF $\geq$ 97\% is so desirable.

   The values of the fitting factors for a couple of template banks against the TaylorEt 3.5PN  waveforms 
are given in Tables~\ref{tab:paramErrTaylorT1Q1}, \ref{tab:paramErrTaylorT1Q1_3}, 
and \ref{tab:paramErrTaylorT1Q1_4}. The
FFs were computed using two separate codes. One of these employs 
LAL \cite{LAL}, 
which is used by the LSC in its inspiral
searches, and the other is a home-grown code
that extensively uses routines from {\it Numerical Recipes} \cite{Num_Rec}.
Both codes have the
ability to compute the FF in Eq. (\ref{FFDef}) as well as the 
more conservative (or lower) {\em minimax} match, detailed in Ref.~\cite{DIS}.
The latter is obtained by 
minimizing the FF of Eq. (\ref{FFDef}) with respect to the coalescence 
phase of the target waveform. 
The numbers presented in the tables are FFs (and not minimax matches)
and, therefore, are larger than values that are realistically achievable
with the above listed inspiral 
template banks, available in LAL,
and the TaylorT4 template bank.

Importantly,
the first few detections will likely require validation from more than
one detector, which implies that in addition to being effectual a template
bank also must be {\it faithful} \cite{DIS}. The latter requirement means that
the parameter values of the best matched template are allowed to differ 
from (a subset of) those of the signal only by acceptably small biases.
This is because unless these systematics are modeled for, the same signal
can be picked up by templates with parameter values different enough 
in two (or more) detectors so as to fail a parameter-value coincidence test
\cite{ethinca}. 
We infer that differences in the estimated masses, illustrated in Tables~\ref{tab:sh_f}
-\ref{tab:paramErrTaylorT1Q1_4},
between different comparable 
class detectors, such as AdLIGO and Virgo, are due to their different
noise PSDs.

Based on the tables and figures presented here, a few observations are in 
order. First, a good fraction of equal-mass compact binary templates, which
are chosen to be from TaylorT1, TaylorT4 (presented only in the figures), 
and TaylorF2 (presented only in the tables) 3.5PN families, 
have ${\rm FF} ~{}^{\Large <}_{\sim}~ 0.97$.
They also show substantial biases for the estimated total-mass
against TaylorEt (3.5PN) signals as long as the symmetric mass-ratio of 
templates is limited to 
$\eta' \leq 0.25$, which is the upper-limit for physical signals.
Note how in the first row of plots in Fig. \ref{fig:T1T4Etq1} the FF first decreases as $m$ is increased before recovering to higher values eventually. This behavior can be explained by the fact that in any given signal band the
TaylorEt approximant has a greater number of GW cycles than the 
$x$-based templates of the same mass system. This means that $x$-based 
templates with $m'<m$ and
$\eta'>\eta$ are more likely to provide a higher match than the
one with $m'=m$ and $\eta'=\eta$. However, since for the equal-mass signals
in Table~\ref{tab:paramErrTaylorT1Q1} and Fig. \ref{fig:T1T4Etq1}
we restrict the templates to have $\eta' \leq 0.25$, the
templates yielding the highest match saturate this bound and attain 
$\eta'= \eta =0.25$. This wave-based
argument alone also implies that the highest match will decrease with increasing $m$. This is 
because while decreasing $m'$ increases the number of template cycles,
which helps in improving the match, decreasing it too much adversely
affects the match by lowering the template $f_{\rm lso}$ and, thereby,
decreasing the integration band. The reason why the FF values eventually 
regain high values at large $m$ is that there the number of wave cycles 
is small and it is easier to obtain larger fits on signals 
with a smaller number of time-frequency bins.

Compact binaries with mass-ratios smaller than unity
can yield high FFs, but at the expense of introducing
high systematic errors in estimating the values of $m$ and $\eta$.
The high FFs can be explained by the fact that unlike in the case of
equal-mass signals, here $\eta'$ can exceed
$\eta$, simply because $\eta<0.25$.
For illustration purposes, we consider signals with two 
different values of the mass-ratio, namely,
$q \equiv m_1/m_2 = 1/3\,\,\mbox{and} \,\, 1/4$. 
In general, our inspiral templates are 
found to be fairly {\it unfaithful} with respect to the fiducial signal
for these cases.
Moreover, Tables~\ref{tab:paramErrTaylorT1Q1} and 
Fig. \ref{fig:ambiguityT1Et10_10Msun}
show that the TaylorT1 template with the maximum match 
almost always has $\eta'=0.25$. This 
arises from restricting the template banks to have physical values of 
the symmetric mass-ratio, namely, $\eta' \leq 0.25$.
As shown in Fig. \ref{fig:ambiguityT1Et10_10Msun}, this 
can be seen from the fact 
that the match in Eq. (\ref{match}), also known as the ambiguity function 
in this context, has a sharp wedge that rises with increasing
$\eta'$ and attains its maximum beyond $\eta'=0.25$. 

The above observation prompted us to compute FFs in 
Tables~\ref{tab:paramErrTaylorT1Q1WideEta} with $\eta' > 0.25$. We find
that it requires $\eta'$ to be as high as 0.35 for the FF to attain 
values at least equal to 97\% for the total-mass range considered here
(i.e., $m \leq 40 M_\odot$), albeit, at the cost of large biases in the 
estimated values
of both $m'$ (up to almost 20\%) and $\eta'$ (up to about 40\%). 
In our opinion, while this allowance increases the
match, it does not necessarily translate into an increase in 
the detection confidence. This is because an expanded range in $\eta'$
has the potential to increase the false-alarm rate. To test what the
effective gain is it is imperative to include templates with $\eta'>0.25$
in signal simulation studies 
involving real interferometric data.

As presented in Tables~\ref{tab:paramErrTaylorT1Q1}-\ref{tab:paramErrTaylorT1Q1_3},
the systematic errors also show variation with respect to the shape of the
detector noise power spectral density. This implies, e.g., that the estimated
value of the total mass of a signal in LIGO and Virgo can disagree and, 
consequently, fail a sufficiently stringent mass-consistency check
in a multi-detector search \cite{Abbott:2007xi,Pai:2000zt}. To wit, in the 
search for inspiral signals in LIGO data from its third and fourth
science runs by the LSC, the estimated chirp-mass, ${\cal M}_c$, in the three 
LIGO detectors was allowed to differ by 0.020$M_\odot$. 
A comparison of the estimated total-mass values in  
Tables~\ref{tab:paramErrTaylorT1Q1} shows that this window needs to be 
relaxed if the search involved both LIGO and Virgo detectors and if 
TaylorEt was indeed a more 
appropriate
representation of a GW signal. 
For instance, Table~\ref{tab:paramErrTaylorT1Q1} shows that 
for $m=5.0~M_\odot$ the measured $m'$ in LIGO and Virgo
are expected to differ by as much as 0.062$M_\odot$, which amounts to
$\Delta{\cal M}_c = 0.026M_\odot$ (where we assumed, conservatively, that 
all the error in ${\cal M}_c$ arose from the error in $m'$). 
This is larger than the allowed 
window and can, therefore, fail a multi-detector mass-consistency test.
The biases shown in Tables~\ref{tab:paramErrTaylorT1Q1} only get worse
as the total mass of the signal is increased. Note that this exercise is meant
to serve as a guide for sources of systematic effects and 
how to deal with them; it is not clear if a concurrent search with LIGO 
and Virgo detectors 
with LIGO and Virgo design sensitivity curves, respectively, (as used for 
Table~\ref{tab:paramErrTaylorT1Q1})
is likely. (Nevertheless, it is more likely that the shapes of the 
respective curves are maintained in the next set of science runs, such as
with the planned Enhanced-LIGO design. This is because the 
FFs depend on the shape 
of these curves and not the overall scale).
It is possible, however, to use our studies to model the
variation of the estimated parameter bias in real detector data 
so that the windows can be scaled and shifted appropriately to mitigate 
the effect on detection efficiency.

\section{Conclusions}
\label{conclusions}
In this paper, we investigated the GW data analysis implications 
of the TaylorEt approximant at the 3.5PN order.
We limited our attention to the case of GW signals from 
non-spinning, comparable mass, compact binaries
in LIGO, AdLIGO and Virgo interferometers.
With the help of detailed fitting factor computations,
we compared the performance of three $x$-based inspiral templates, namely, 
TaylorT1, TaylorT4, and TaylorF2
at 3.5PN order,
in detecting a fiducial TaylorEt signal of the same PN order.
For the equal-mass binaries, we generally obtain FF ${}^{\Large <}_{\sim}~ 0.97$ 
when restricting the above templates to physically allowed mass-ratios.
In the case of unequal mass binaries, it is possible to obtain high FFs with the
LAL inspiral templates.
However, the templates that provide those high FFs have substantially different values of
 $m$ and $\eta$ compared to those of the fiducial TaylorEt signals. 
In all cases, templates giving high FFs have lower values of the total-mass parameter compared to 
their associated TaylorEt signals. This is due to the fact that in a given GW frequency band
the TaylorEt approximant always provides more accumulated GW cycles than the 
$x$-based templates. 
Further, the systematic errors in $m$ and $\eta$ parameters of TaylorF2 templates are substantially
higher than the statistical errors in those parameters reported in Ref.~\cite{AISS}.
These observations lead us to believe that the unfaithful nature of the $x$-based inspiral templates vis \`{a} vis the TaylorEt approximant
may adversely affect the chances of detecting GW inspiral signals assuming
that the latter waveforms more accurately represent such a signal.

To summarize, the present study shows that it should be worthwhile
to include the theoretically motivated TaylorEt templates,
which has a number of attractive features as detailed in Ref.~\cite{AG07},
in the search for inspiral GW signals from non-spinning compact
binaries in the data of ground-based broadband detectors.
Further, this work should also be useful in assessing the effects of
systematic errors arising from employing inspiral templates from
different representations on a GW signal's detectability and
parameter estimation with earth-based detectors.

  In the literature, there exist gauge-dependent prescriptions for constructing inspiral $h(t)$ that give equal emphasis to both the conservative and the reactive orbital phase
evolution, such as the Effective One-Body (EOB) approach \cite{DamourEOB} and 
the Semi-Analytic Puncture Evolution
(SAPE) \cite{SAPE}. Recall that the conservative Hamiltonian
relevant for the EOB scheme is in Schwarzschild-type coordinates, while for SAPE it is in the Arnowitt-Deser-Misner gauge.
By contrast, PN-accurate TaylorEt based $h(t)$ is fully gauge-invariant. Therefore, we
are pursuing a study, similar to the one presented here,
of comparing the effectualness and faithfulness of 
EOB- and SAPE-based inspiral waveforms vis \'{a} vis
TaylorEt waveforms. We are also extending the present analysis
by including spin effects with the help of a generalized version of 
fiducial signals and templates detailed in Ref.~\cite{HHBG}.

\acknowledgments
It is our pleasure to thank Bruce Allen and
Gerhard Sch\"afer for helpful discussions and persistent
encouragement. We are grateful for their warm hospitality at 
Hannover and Jena during various stages of the present study.
We would also like to thank B. S. Sathyaprakash for useful
observations on the comparison of the different template banks 
studied here and G.~Esposito-Far\`ese and 
C.~R\"{o}ver for their careful reading of the
manuscript and making helpful suggestions.
SB thanks P.~Ajith for his explanation of a fast algorithm for
computing the fitting
factor of a generic template bank and a target signal, and for
providing useful comments on this work.
This work is supported in part by
the NSF grant PHY-0239735 to SB  and by grants from
DLR (Deutsches Zentrum f\"ur Luft- und Raumfahrt)
and DFG's SFB/TR 7 ``Gravitational Wave Astronomy'' to AG and MT,
respectively.

\begin{widetext}

\newpage
\begin{table}[!ht]
\caption{
The analytic fits to the one-sided noise power-spectral densities, $S_n(f)$,
of LIGO, Virgo and AdLIGO employed in this paper. The expressions for $S_n(f)$ are expressed in terms of 
$y=f/f_0$, where the ``knee frequency'' $f_0$ takes values of $150$Hz, $500$Hz and 
$215$Hz for the LIGO, Virgo and AdLIGO, respectively. Notice that  $S_n(f)$  rises 
sharply above the seismic cut-off $f_s$. We drop an overall scale factor from each of the expressions of $S_n(f)$ below as it does not affect our fitting factor studies.
}
\vskip 5pt
\begin{tabular}{|c|c|c|}
\hline 
Detector ~~~&~~~ $f_s$ ~~~&~~~ $S_n(f)$ (up to an overall scale) \\
\hline 
LIGO ~~~&~~~ 40 Hz ~~~&~~~  $ \biggl \{   
(4.49\,y)^{-56} + 0.16\, {y}^{-4.52} + 0.52 + 0.32\, {y}^2
 \biggr \} $\\
\hline 
Virgo ~~~&~~~ 20 Hz ~~~&~~~ $ \biggl \{   (6.23 y)^{-5} + 2 y^{-1} + 1 + y^ 2 \biggr \}  $ \\
\hline 
AdLIGO ~~~&~~~ 20 Hz ~~~&~~~ $ \biggl \{  y^{-4.14} - 5y^{-2} + \frac {111 (1-y^2+y^4/2)}{(1+y^2/2)}
\biggr \} $\\
\hline
\end{tabular}
\label{tab:sh_f}
\end{table}

\begin{table}[!ht]
\caption{
Values of the fitting factors and the associated systematic errors in the total mass parameter 
relevant 
for the LIGO, Virgo and AdLIGO detectors
while employing 
the TaylorEt approximant at 3.5PN order to model the fiducial inspiral signal.
The search templates, extracted from LAL,
belong to the TaylorT1 approximant at 3.5PN order
in all cases except when the ``Detector'' is denoted as ``AdLIGO-F2''; in the latter
case the templates are given by the TaylorF2 approximant at 3.5PN order. 
The total mass and symmetric mass-ratio parameters of the signal and templates are
denoted by $(m, \eta)$ and $(m',\eta')$, respectively.
We list $m'$ of the search template that gives the largest match for a given
signal. In all the systems below, $\eta'$ was found to be 0.250 and is, therefore, not listed. The fiducial GW signal and search templates are terminated  
when the instantaneous GW frequency reaches the value corresponding to 
the last stable orbit.
The low FFs reported in certain cases may be attributed to the fact that we have not allowed 
$\eta'$ to go beyond its realistic bound of $0.25$. 
The maximum overlap is always for values of $m' < m$.  
The numbers below are for the equal-mass case: $q =m_1/m_2=1$, i.e., $\eta = 0.25$. It is interesting to note that for small $m$, Virgo has the worst FF. This owes its cause to a flatter
noise PSD at low frequencies which weights the phase difference between
the templates and the signal the most there. Also, the FF values 
of TaylorF2 templates are larger than those of TaylorT1 (computed
for comparison only for AdLIGO). This is consistent with the
observation in Fig. \ref{fig:T1T4Etq1} where the TaylorT4 FF values are
also higher than 
those of TaylorT1. (Recall our speculation that TaylorF2 approximants
are essentially stationary phase approximations of TaylorT4
approximants.) 
}
\vskip 5pt
\begin{tabular}{|c|c|c| c |}
\hline ~~~
$m_1$ ($M_\odot$) $\equiv m_2$ ($M_\odot$) ~~~&~~~ Detector ~~~&~~~ FF ~~~&~~~ $m'$ ($M_\odot$) ~~~\\
\hline ~~
1.4     ~~~&~~~ AdLIGO-T1 ~~~&~~~ 0.96 ~~~&~~~ 2.7995 ~~~\\
        ~~~&~~~ AdLIGO-F2 ~~~&~~~ 0.98  ~~~&~~~ 2.8000 ~~~\\
        ~~~&~~~ LIGO   ~~~&~~~ 0.95  ~~~&~~~  2.7985 ~~~\\
        ~~~&~~~ Virgo  ~~~&~~~ 0.93 ~~~&~~~ 2.7999 ~~~\\
\hline
3.0     ~~~&~~~ AdLIGO-T1 ~~~&~~~ 0.90 ~~~&~~~ 5.9924 ~~~\\
        ~~~&~~~ AdLIGO-F2   ~~~&~~~ 0.94 ~~~&~~~ 5.9952  ~~~\\
        ~~~&~~~ LIGO   ~~~&~~~ 0.94 ~~~&~~~ 5.9841 ~~~\\
        ~~~&~~~ Virgo  ~~~&~~~ 0.87 ~~~&~~~ 5.9950 ~~~\\
\hline
5.0     ~~~&~~~ AdLIGO-T1 ~~~&~~~ 0.87 ~~~&~~~ 9.9365 ~~~\\
        ~~~&~~~ AdLIGO-F2 ~~~&~~~ 0.92 ~~~&~~~ 9.9404  ~~~\\
        ~~~&~~~ LIGO   ~~~&~~~ 0.94 ~~~&~~~ 9.9117 ~~~\\
        ~~~&~~~ Virgo  ~~~&~~~ 0.84 ~~~&~~~ 9.9740 ~~~\\
\hline
8.0     ~~~&~~~ AdLIGO-T1 ~~~&~~~ 0.87 ~~~&~~~ 15.7649 ~~~\\
        ~~~&~~~ AdLIGO-F2 ~~~&~~~ 0.92 ~~~&~~~ 15.7624 ~~~\\
        ~~~&~~~ LIGO   ~~~&~~~ 0.95 ~~~&~~~ 15.6663 ~~~\\
        ~~~&~~~ Virgo  ~~~&~~~ 0.86 ~~~&~~~ 15.8770 ~~~\\
\hline
10.0    ~~~&~~~ AdLIGO-T1 ~~~&~~~ 0.89 ~~~&~~~ 19.5718 ~~~\\
        ~~~&~~~ AdLIGO-F2   ~~~&~~~ 0.90 ~~~&~~~ 19.6786 ~~~\\
        ~~~&~~~ LIGO   ~~~&~~~ 0.96 ~~~&~~~ 19.3438 ~~~\\
        ~~~&~~~ Virgo  ~~~&~~~ 0.89 ~~~&~~~ 19.7728 ~~~\\
\hline
15.0    ~~~&~~~ AdLIGO-T1 ~~~&~~~ 0.90 ~~~&~~~ 28.8432 ~~~\\
        ~~~&~~~ AdLIGO-F2   ~~~&~~~ 0.93 ~~~&~~~ 29.4024 ~~~\\
        ~~~&~~~ LIGO   ~~~&~~~ 0.98 ~~~&~~~ 28.3201 ~~~\\
        ~~~&~~~ Virgo  ~~~&~~~ 0.92 ~~~&~~~ 29.3476 ~~~\\
\hline
20.0    ~~~&~~~ AdLIGO-T1 ~~~&~~~ 0.92 ~~~&~~~ 37.9325 ~~~\\
        ~~~&~~~ AdLIGO-F2 ~~~&~~~ 0.89 ~~~&~~~ 37.5663 ~~~\\
        ~~~&~~~ LIGO   ~~~&~~~ 0.98 ~~~&~~~ 36.8851 ~~~\\
        ~~~&~~~ Virgo  ~~~&~~~ 0.93 ~~~&~~~ 38.6133 ~~~\\
\hline
\end{tabular}
\label{tab:paramErrTaylorT1Q1}
\end{table}
\begin{table}[!hbt]
\caption{
Fitting factor values and inherent parameter biases relevant for the LIGO, Virgo and AdLIGO interferometers, for several compact binaries with mass-ratio $q=1/3$ (i.e., $\eta = 0.1875)$). The other details are as in 
Table~\ref{tab:paramErrTaylorT1Q1}.
We observe that high FFs are always for templates characterized by $m'<m$ and $\eta'\simeq 0.25$.
It is clear that the best matched templates are highly biased with respect to their fiducial signals. 
%
}
\vskip 5pt
\begin{tabular}{|c|c|c|c|c|c|}
\hline ~~~
$m_1$ ($M_\odot$) - $m_2$ ($M_\odot$) ~~~&~~~ $m$ ($M_\odot$)   ~~~&~~~ Detector ~~~&~~~ FF ~~~&~~~ $m'$ ($M_\odot$) ~~~&~~~ $\eta'$~~~\\
\hline ~~~
3-9	~~~&~~~ 12	~~~&~~~ AdLIGO-T1  ~~~&~~~ 0.98	~~~&~~~ 10.1630 ~~~&~~~ 0.250 ~~~\\
        ~~~&~~~       ~~~&~~~ AdLIGO-F2  ~~~&~~~ 0.97     ~~~&~~~ 10.2269 ~~~&~~~ 0.248 ~~~\\
	~~~&~~~ 	~~~&~~~ LIGO    ~~~&~~~ 0.98  ~~~&~~~ 10.1588 ~~~&~~~ 0.250 ~~~\\
	~~~&~~~ 	~~~&~~~ Virgo   ~~~&~~~ 0.97  ~~~&~~~ 10.1596 ~~~&~~~ 0.250 ~~~\\
\hline 
4-12	~~~&~~~ 16	~~~&~~~ AdLIGO-T1  ~~~&~~~ 0.97	~~~&~~~ 13.5519 ~~~&~~~ 0.250 ~~~\\
        ~~~&~~~       ~~~&~~~ AdLIGO-F2  ~~~&~~~ 0.98     ~~~&~~~ 13.5704 ~~~&~~~ 0.250 ~~~\\
	~~~&~~~ 	~~~&~~~ LIGO    ~~~&~~~ 0.98  ~~~&~~~ 13.5078 ~~~&~~~ 0.250 ~~~\\
	~~~&~~~ 	~~~&~~~ Virgo   ~~~&~~~ 0.97  ~~~&~~~ 13.5522 ~~~&~~~ 0.250 ~~~\\
\hline 
5-15	~~~&~~~ 20	~~~&~~~ AdLIGO-T1  ~~~&~~~ 0.96	~~~&~~~ 16.9127 ~~~&~~~ 0.250 ~~~\\
        ~~~&~~~       ~~~&~~~ AdLIGO-F2   ~~~&~~~  0.97   ~~~&~~~ 16.9609 ~~~&~~~ 0.250 ~~~\\
	~~~&~~~ 	~~~&~~~ LIGO    ~~~&~~~ 0.98  ~~~&~~~ 16.8320 ~~~&~~~ 0.250 ~~~\\
	~~~&~~~ 	~~~&~~~ Virgo   ~~~&~~~ 0.97  ~~~&~~~ 16.9299 ~~~&~~~ 0.250 ~~~\\
\hline 
7-21	~~~&~~~ 28	~~~&~~~ AdLIGO-T1  ~~~&~~~ 0.96	~~~&~~~ 23.6070 ~~~&~~~ 0.250 ~~~\\
        ~~~&~~~       ~~~&~~~ AdLIGO-F2  ~~~&~~~ 0.96     ~~~&~~~ 23.7107 ~~~&~~~ 0.250 ~~~\\
	~~~&~~~ 	~~~&~~~ LIGO    ~~~&~~~ 0.99  ~~~&~~~ 23.4532 ~~~&~~~ 0.247 ~~~\\
	~~~&~~~ 	~~~&~~~ Virgo   ~~~&~~~ 0.97  ~~~&~~~ 23.6543 ~~~&~~~ 0.250 ~~~\\
\hline 
10-30	~~~&~~~ 40	~~~&~~~ AdLIGO-T1  ~~~&~~~ 0.98	~~~&~~~ 33.5956 ~~~&~~~ 0.250 ~~~\\	
        ~~~&~~~       ~~~&~~~ AdLIGO-F2  ~~~&~~~ 0.95     ~~~&~~~ 33.8244 ~~~&~~~ 0.249 ~~~\\
	~~~&~~~ 	~~~&~~~ LIGO    ~~~&~~~ 0.98  ~~~&~~~  35.4216 ~~~&~~~ 0.208 ~~~\\
	~~~&~~~       ~~~&~~~ Virgo   ~~~&~~~ 0.98  ~~~&~~~ 33.6106 ~~~&~~~ 0.250 ~~~\\
\hline
\end{tabular}
\label{tab:paramErrTaylorT1Q1_3}
\end{table}

\newpage

\begin{table}[!hbt]
\caption{
Fitting factor values, relevant for the LIGO, Virgo, and AdLIGO, for several compact binaries
having mass ratio $q=1/4$ or $\eta =0.16$. The other details are as in Table~\ref{tab:paramErrTaylorT1Q1}, 
and our conclusions are also similar to those reported in Table~\ref{tab:paramErrTaylorT1Q1_3}. Additionally, it is interesting to observe that across both Table~\ref{tab:paramErrTaylorT1Q1_3} and the current table, as the total mass of the system is increased in LIGO, the bias in  $\eta'$ there appears to decrease noticably. This is because LIGO has a higher $f_s$ than the other detectors considered here 
(see Table~\ref{tab:sh_f}). Correspondingly, the number of in-band cycles is considerably reduced in it for higher mass, to the extent that tuning only one parameter, $m'$ , in LIGO suffices to attain higher FFs.
}
\vskip 5pt
\begin{tabular}{|c|c|c|c|c|c|}
\hline ~~~
$m_1$ ($M_\odot$) - $m_2$ ($M_\odot$) ~~~&~~~ $m$ ($M_\odot$)   ~~~&~~~ Detector ~~~&~~~ FF ~~~&~~~ $m'$ ($M_\odot$) ~~~&~~~ $\eta'$ ~~~\\
\hline ~~~
3-12	~~~&~~~ 15	~~~&~~~ AdLIGO-T1 ~~~&~~~ 0.97	~~~&~~~ 11.7544	~~~&~~~ 0.246 ~~~\\
        ~~~&~~~       ~~~&~~~ AdLIGO-F2   ~~~&~~~  0.97      ~~~&~~~  12.2512     ~~~&~~~ 0.228 ~~~\\
        ~~~&~~~       ~~~&~~~ LIGO   ~~~&~~~ 0.99  ~~~&~~~ 11.9641  ~~~&~~~ 0.238 ~~~\\
        ~~~&~~~       ~~~&~~~ Virgo   ~~~&~~~ 0.95  ~~~&~~~ 11.7448  ~~~&~~~ 0.246 ~~~\\
\hline 
4-16	~~~&~~~ 20	~~~&~~~ AdLIGO-T1 ~~~&~~~ 0.98	~~~&~~~ 15.5198	~~~&~~~ 0.250 ~~~\\
        ~~~&~~~       ~~~&~~~ AdLIGO-F2   ~~~&~~~  0.97   ~~~&~~~ 16.0009     ~~~&~~~ 0.237 ~~~\\
        ~~~&~~~       ~~~&~~~ LIGO   ~~~&~~~ 0.99 ~~~&~~~ 15.5188 ~~~&~~~ 0.250 ~~~\\
        ~~~&~~~       ~~~&~~~ Virgo   ~~~&~~~ 0.98  ~~~&~~~ 15.5213  ~~~&~~~ 0.250 ~~~\\
\hline 
5-20	~~~&~~~ 25	~~~&~~~ AdLIGO-T1 ~~~&~~~ 0.98	~~~&~~~ 19.4193	~~~&~~~ 0.250 ~~~\\
        ~~~&~~~       ~~~&~~~ AdLIGO-F2   ~~~&~~~ 0.96    ~~~&~~~  20.2636    ~~~&~~~ 0.231 ~~~\\
        ~~~&~~~       ~~~&~~~ LIGO   ~~~&~~~ 0.99  ~~~&~~~ 20.3615  ~~~&~~~ 0.224 ~~~\\
        ~~~&~~~       ~~~&~~~ Virgo   ~~~&~~~ 0.99  ~~~&~~~ 19.4180  ~~~&~~~ 0.250 ~~~\\
\hline 
28-7	~~~&~~~ 35	~~~&~~~ AdLIGO-T1 ~~~&~~~ 0.96	~~~&~~~ 27.2169	~~~&~~~ 0.250 ~~~\\
        ~~~&~~~       ~~~&~~~ AdLIGO-F2   ~~~&~~~  0.96    ~~~&~~~  27.6552     ~~~&~~~ 0.243 ~~~\\
        ~~~&~~~       ~~~&~~~ LIGO   ~~~&~~~ 0.98  ~~~&~~~ 28.7005  ~~~&~~~ 0.217 ~~~\\
        ~~~&~~~       ~~~&~~~ Virgo   ~~~&~~~ 0.98  ~~~&~~~ 27.2233  ~~~&~~~ 0.250 ~~~\\
\hline 
8-32	~~~&~~~ 40	~~~&~~~ AdLIGO-T1 ~~~&~~~ 0.98	~~~&~~~ 32.0353	~~~&~~~ 0.234 ~~~\\
        ~~~&~~~       ~~~&~~~ AdLIGO-F2   ~~~&~~~  0.95      ~~~&~~~  32.0962     ~~~&~~~ 0.235 ~~~\\
        ~~~&~~~       ~~~&~~~ LIGO   ~~~&~~~ 0.98  ~~~&~~~ 35.9969 ~~~&~~~ 0.167 ~~~\\
        ~~~&~~~       ~~~&~~~ Virgo   ~~~&~~~ 0.98  ~~~&~~~ 31.9093  ~~~&~~~ 0.236 ~~~\\
\hline
\end{tabular}
\label{tab:paramErrTaylorT1Q1_4}
\end{table}

\begin{table}[!hbt]
\caption{
The recomputed FF values for AdLIGO for compact binaries 
studied in Table~\ref{tab:paramErrTaylorT1Q1}. Unlike in that table,
here we allowed the symmetric mass-ratio parameter of the templates to
vary beyond the physically allowed range,
and up to $\eta'=0.35$. As shown below, doing so improves the FFs to be higher
than or equal to the desirable lower-limit of 97\%. However, this gain comes
at the cost of high biases in estimated masses and, possibly, an increased
noise contribution to the detection statistic.
}
\vskip 5pt
\begin{tabular}{|c|c|c|c|}
\hline ~~~
$m_1$ ($M_\odot$)  ~~~&~~~ FF ~~~&~~~ $m'$ ($M_\odot$) ~~~&~~~ $\eta'$ ~~~\\
\hline ~~~
1.4  ~~~&~~~  0.99 ~~~&~~~    2.7350     ~~~&~~~  0.260  ~~~\\
3    ~~~&~~~  0.99 ~~~&~~~    5.5939     ~~~&~~~  0.288  ~~~\\
5    ~~~&~~~  0.99 ~~~&~~~    8.8575     ~~~&~~~  0.308  ~~~\\
8    ~~~&~~~  0.99 ~~~&~~~   13.4103     ~~~&~~~  0.341  ~~~\\
10   ~~~&~~~  0.98 ~~~&~~~   16.5826     ~~~&~~~  0.348  ~~~\\
15   ~~~&~~~  0.97 ~~~&~~~   24.7678     ~~~&~~~  0.350  ~~~\\
20   ~~~&~~~  0.97 ~~~&~~~   32.6322     ~~~&~~~  0.350  ~~~\\
\hline
\end{tabular}
\label{tab:paramErrTaylorT1Q1WideEta}
\end{table}


\begin{figure}[!ht]
\includegraphics[height=8.5cm, width=8.cm]{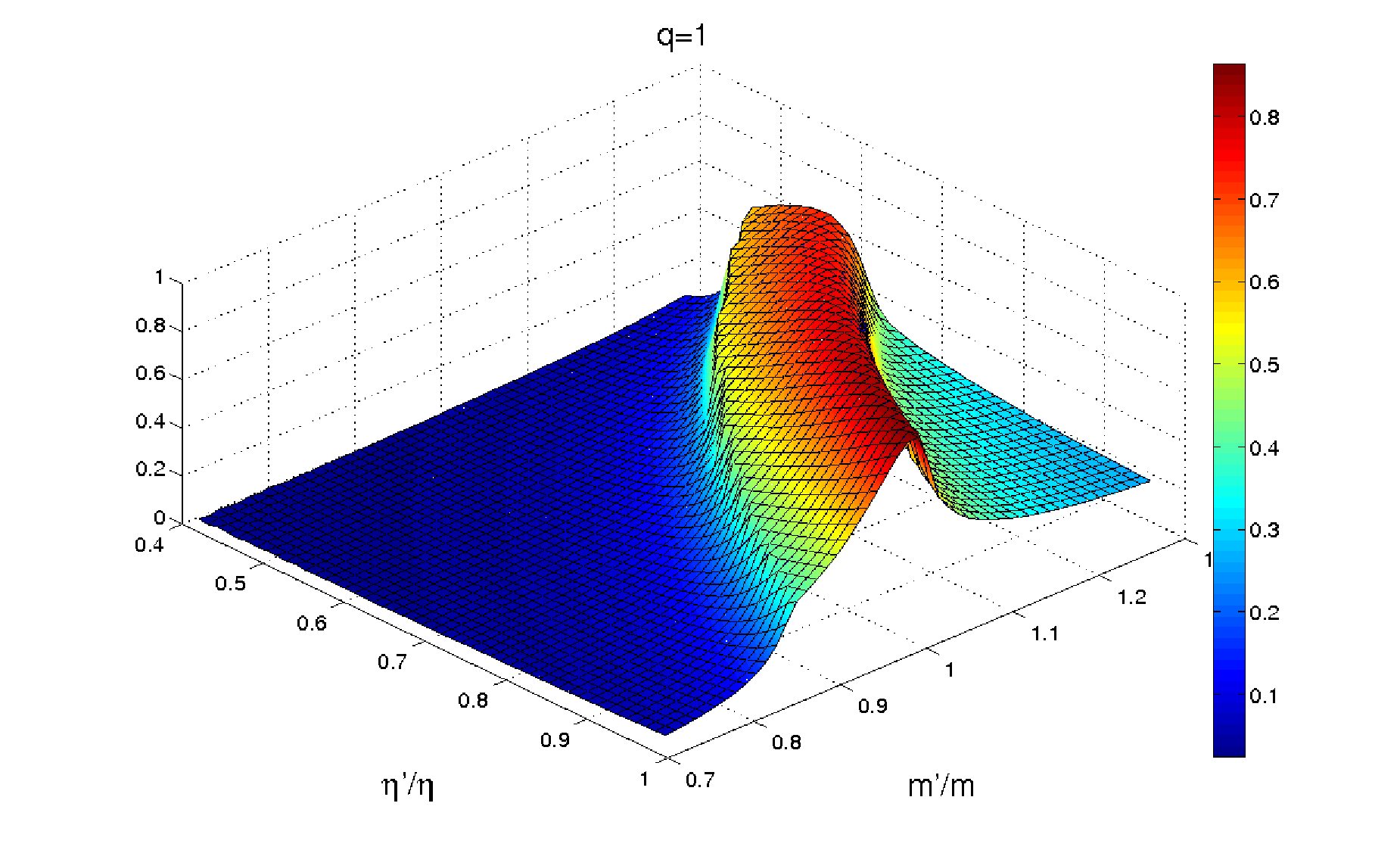}\\[0.1cm]
\includegraphics[height=8.5cm, width=8.cm ]{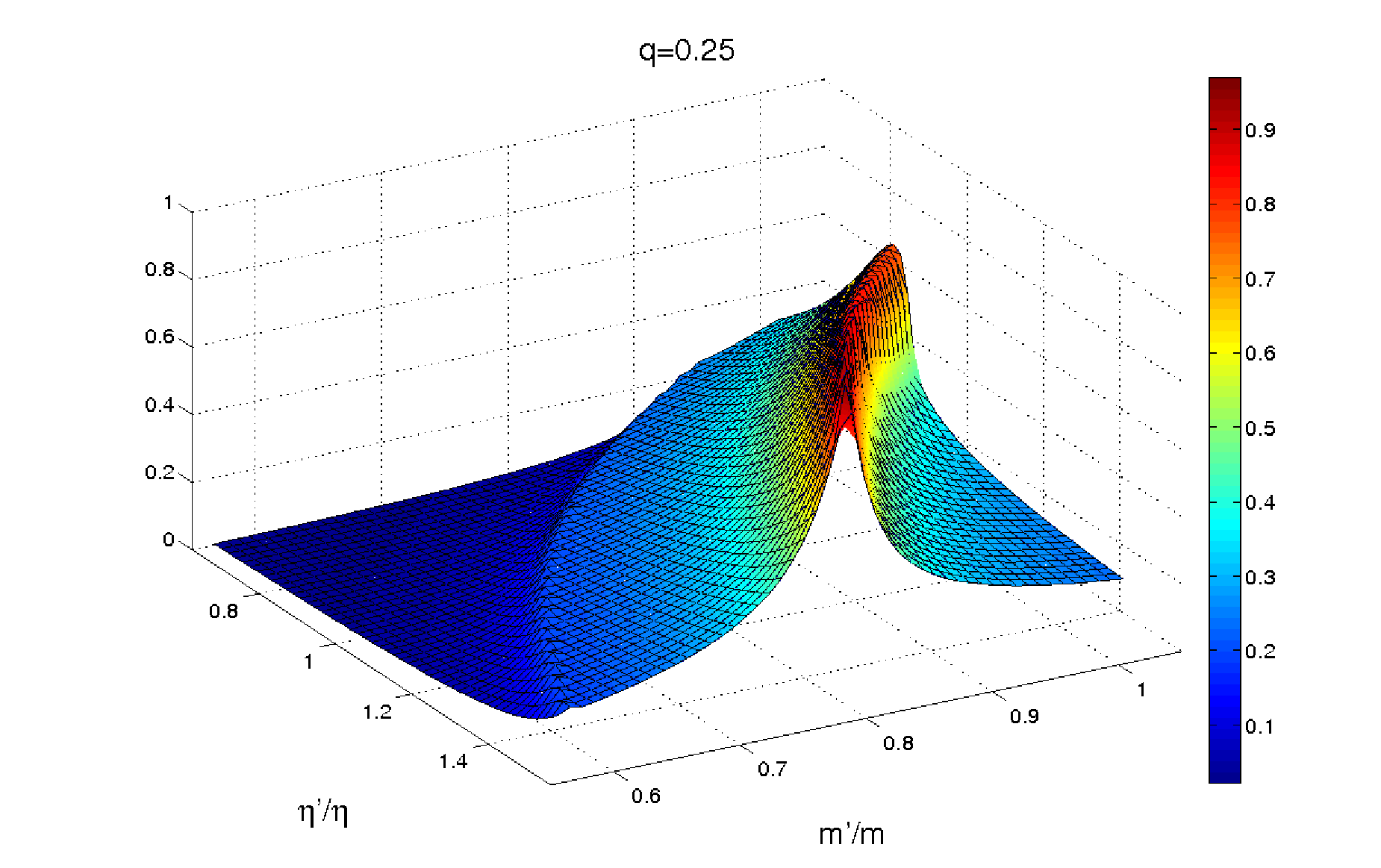}
\caption{
The plots display the 
match $M$ (see Eq. \ref{match})
\cite{DIS}
as a function of the parameters $(m',\eta')$
of the search templates.
The plot axes are $\eta'/\eta$ and $m'/m$, where $(m,\eta)$ are the 
values of the TaylorEt signal parameters.
For the equal-mass case ($q=1$), the match keeps rising as $\eta'$ is increased
and attains its global maximum in the unphysical region $\eta' > 0.25$.
In the $q=1/4$ plot, 
the maximum value of $M$ is around $\eta' \sim 0.25$, which is
much larger than the signal parameter value of 
$\eta=0.16$. 
The maximum match remains fairly high, at more than 0.95, on
certain crests of the ridges. 
}
\label{fig:ambiguityT1Et10_10Msun}
\end{figure}

\newpage
\begin{figure}[!ht]
\includegraphics[height=7.0cm, width=14.0cm]{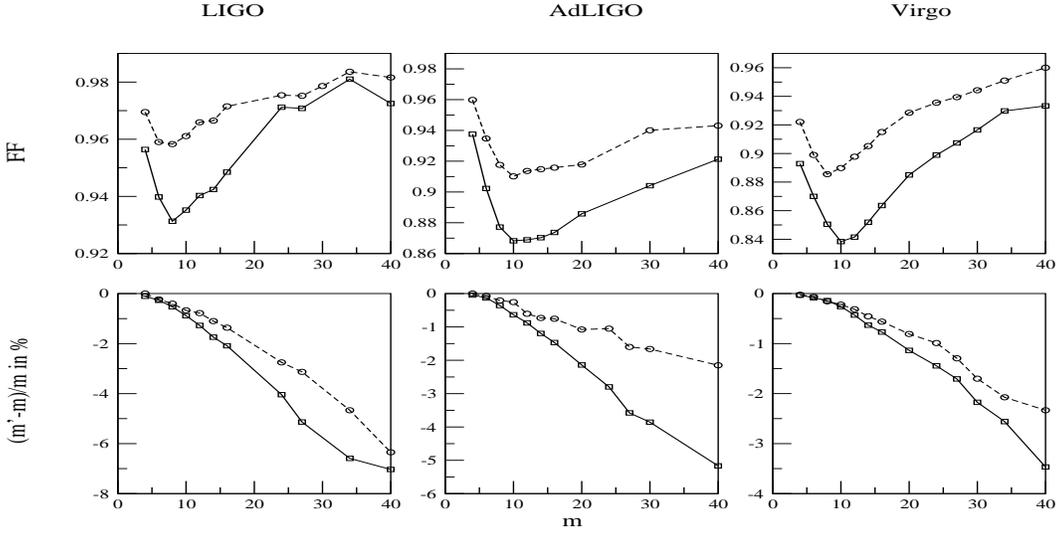}
\caption{
A collection of plots that summarizes a set of results from our fitting-factor studies involving the TaylorEt, TaylorT1/TaylorT4 approximants 
at 3.5PN order for equal-mass binaries.
The first row provides plots of FF against the total mass of the inspiral signals and the second
row shows fractional error (in percentage) in the estimated total-mass values as a function of
the signal's total-mass.
The thick and dashed lines denote results for the TaylorT1 and 
TaylorT4 approximants, respectively.
For high mass binaries, we observe high biases in the $m'$ values. Above, we do not provide
$\eta'$ versus $m$ plots as the best matched filters always have $\eta' \simeq 0.25$.
}
\label{fig:T1T4Etq1}
\end{figure}

\newpage

\begin{figure}[!ht]
\includegraphics[height=7cm, width=15cm]{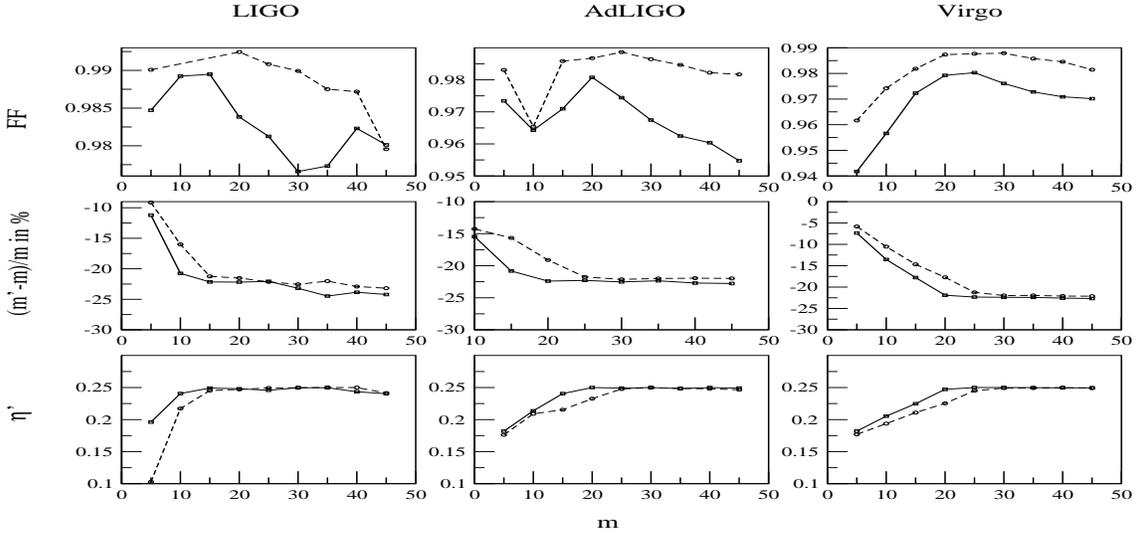}\\[0.1cm]
\caption{
A set of plots summarizing our fitting-factor and parameter estimation results
for the TaylorEt, TaylorT1 approximants at 3.5PN order 
for compact binaries having $q=1/4$.
The first two rows show plots analogous to those presented in 
Fig.~\ref{fig:T1T4Etq1}.
The third row shows plots of $\eta'$ versus $m$. (Note that
$\eta = 0.16$ in all these plots.)
Here too the thick and dashed lines are for the TaylorT1 and
TaylorT4 approximants, respectively.
We clearly observe substantially higher biases for $\eta'$ that for
$m'$ values and in most cases, the best matched templates always have $\eta' \simeq 0.25$.
}
\label{fig:T1Etq14}
\end{figure}

\end{widetext}

\begin{thebibliography}{999}
\bibitem{ELIGOFFs}
Since the shape of the LIGO noise PSD is expected to remain mostly 
unchanged in Enhanced LIGO, we expect the FF numbers and mass-error 
estimates given here to be roughly applicable for that detector as well.
\end{thebibliography}

\end{document}